\documentstyle[12pt,bezier]{article}

\newcommand{\bra}[1]{\langle #1 |}
\newcommand{\ket}[1]{| #1 \rangle}

\newcommand{\bm}[1]{\mbox{\boldmath $#1$}}

\def\e{\mbox{e}}

\begin{document}

\begin{titlepage}
\hbox to \hsize{\hfil hep-th/9510107}
\hbox to \hsize{\hfil INR 15/10/95}
\hbox to \hsize{\hfil October, 1995}
\vfill
\large \bf
\begin{center}
STRINGS AS A MODEL FOR PARENT AND BABY \\
UNIVERSES: TOTAL SPLITTING RATES
\end{center}
\vskip1cm
\normalsize
\begin{center}
{\bf Kh. S. Nirov\footnote{E--mail: nirov@ms2.inr.ac.ru}~~~ and~~~
      V. A. Rubakov\footnote{E--mail: rubakov@ms2.inr.ac.ru}}\\
{\small \em Institute for Nuclear Research of the Russian Academy of
Sciences,} \\
{\small \em 60th October Anniversary prospect 7a, Moscow 117312}
\end{center}
\vskip2cm
\begin{abstract}
\noindent
Emission of hard microscopic string (graviton) by an excited
macroscopic string may be viewed as a model of branching of
a $(1+1)$-dimensional baby universe off large parent one.
We show that, apart from a trivial factor, the total emission
rate is not suppressed by the size of the macroscopic string.
This implies unsuppressed loss of quantum coherence in
$(1+1)$-dimensional parent universe.
\end{abstract}
\vfill
\end{titlepage}

{\bf 1.} Theory of fundamental strings in critical dimension may serve
as a model for parent and baby universes \cite{Hawking2,LyHa,VR}.
Long smooth strings carrying particle-like excitations may be viewed as
large $(1+1)d$ universes, while microscopic string states (gravitons
and alike) model baby universes. Within this model, one may try to
understand various issues which were originally discussed in the context
of $(3+1)$-dimensional  theory of gravity
\cite{Hawking1,LRT,GS1,Coleman1,GS2,KSB,Banks} (for further motivation
see ref. \cite{VR}, which we will refer to as I in this paper).
In particular, it has been argued in I that the emission of a baby
universe (hard graviton) into $D$-dimensional target space-time,
induced by ``particles'' in the large universe (macroscopic string),
leads to the loss of
quantum coherence for $(1+1)d$ observer living in the parent universe.
It is then of interest to estimate the dependence of the corresponding
emission rate on the size of the parent universe, $L$.
{}From $D$-dimensional point of view this is the rate of the decay of
a slightly excited  macroscopic string into a graviton and another
macroscopic string. The results of I concerning this rate were not
decisive: only special final states of the macroscopic string were
considered, and the partial decay rates into these particular final
states were suppressed at large $L$, albeit only logarithmically.

In this paper we show that the {\em total} rate of the emission of hard
gravitons by excited macroscopic string is unsuppressed at large $L$
(apart from a trivial factor). In $(1+1)d$ language this means that the
branching off of  baby universes, induced by interactions of $(1+1)d$
``particles'', is finite for parent universes of large size. As
argued in I, this in turn implies that the loss of quantum coherence
in the large $(1+1)d$ universe occurs at finite, $L$-independent rate.

\vskip3mm

{\bf 2.} The construction of excited macroscopic states of bosonic closed
string in critical dimension, outlined in I, is as follows. One considers
$D$-dimensional target space with one dimension, $X^1$, compactified to
a large circle of length $2\pi L$. Let $\ket{{\cal P}}$
be the ground state
of the string winding once around this compact dimension. In its rest
frame
\begin{equation}
 {\cal P}=(M_{0},{\bf 0})
\label{4*}
\end{equation}
with
\[
    M_{0}^{2}=4L^{2} - 8
\]
(we follow conventions of ref. \cite{GSW}). The excited string states
are constructed by making use of the DDF operators
\[
   a^{\alpha}_{n}=\int\limits_{0}^{\pi}~\frac{d\sigma_{+}}{\pi}
              \exp\left[4in\frac{e_{\mu}
          X^{\mu}_{L}(\sigma_{+})}{e_{\mu}P^{\mu}_{L}}\right]
             \xi^{\alpha}_{i}\partial_{+}X^{i}_{L}(\sigma_{+}) \,,
\]
\[
   \tilde{a}^{\alpha}_{\tilde{n}}
   =\int\limits_{0}^{\pi}~\frac{d\sigma_{-}}{\pi}
              \exp\left[4i\tilde{n}\frac{e_{\mu}
          X^{\mu}_{R}(\sigma_{-})}{e_{\mu}P^{\mu}_{R}}\right]
             \xi^{\alpha}_{i}\partial_{-}X^{i}_{R}(\sigma_{-}) \,,
\]
where $e^{\mu}$ is an arbitrary (but fixed) light-like vector in $D$
dimensions with $e^0 = 1$, ${\bm \xi}^{\alpha}$ ($\alpha = 1,\dots,D-2$)
are orthonormal spatial vectors which are orthogonal to ${\bm e}$,
\[
   P^{\mu}_{L}= P^{\mu} + 2L^{\mu}\,, ~~~~~~
   P^{\mu}_{R}= P^{\mu} - 2L^{\mu}
\]
with $L^{\mu} = (0,L,0,\dots,0)$, and $X^{\mu}_{L,R}$ are usual
left and right components of the string coordinate operator in the
sector of winding string \cite{GSW}, e.g.,
\[
   X^{\mu}_{L}(\sigma_{+})=
         \frac{1}{2}X^{\mu} + \frac{1}{2}P^{\mu}_{L}\sigma_{+} +
         \frac{i}{2}\sum_{k\neq 0}\frac{1}{k}\alpha^{\mu}_{k}
         \mbox{e}^{-2ik\sigma_{+}} \,.
\]
The macroscopic string state with two particle-like excitations is
\begin{equation}
    \ket{n,\alpha;\tilde{n},\beta} =
	\frac{1}{\sqrt{n\tilde{n}}}a^{\alpha}_{-n}
        \tilde{a}^{\beta}_{-\tilde{n}} \ket{{\cal P}} \,,
\label{5*}
\end{equation}
where normalization corresponds to ``one particle per volume $L$'' in
$(1+1)$ dimensions. The state (\ref{5*}) obeys the Virasoro constraints
provided that
\[
  \frac{n}{(e{\cal P}_{L})}=
  \frac{\tilde{n}}{(e{\cal P}_{R})} =
  \frac{n+\tilde{n}}{2(e{\cal P})} \,.
\]
Its target space momentum and mass are
\begin{equation}
    P^{\mu} = {\cal P}^{\mu} - \frac{2(n+\tilde{n})}{(e{\cal P})} e^{\mu} \,,
\label{5+}
\end{equation}
\[
     M^{2}=-P^{2}=M_{0}^{2} + 4(n+\tilde{n}) \,.
\]
Roughly speaking, the state (\ref{5*}) contains one particle moving
right in $(1+1)$-di\-men\-si\-on\-al universe and one particle moving
left, the bare $(1+1)d$ momenta of these particles being $n/L$ and
$(-\tilde{n}/L)$, respectively{\footnote {Of course, this
 interpretation should not be taken literally, as we are dealing with
conformal field theory in $(1+1)$ dimensions.  Nevertheless, we will
use somewhat loose notion of ``particles'' in what follows, since the
analogy to usual particles is close enough.}}. Therefore, we will be
interested in the regime
\begin{equation}
  L\to \infty,~~~~\frac{n}{L},~\frac{\tilde{n}}{L} = fixed, \label{6*}
\end{equation}
which corresponds to  large universe with ``particles''
of finite $(1+1)d$ bare momenta.

\vskip3mm

{\bf 3.} The  main  purpose of this paper is to evaluate the $L$-dependence,
in the regime (\ref{6*}), of the total decay rate of the state
(\ref{5*}) into another excited macroscopic string state and one graviton
whose target space momentum is finite at large $L$  (in this sense the
graviton is hard). This rate is proportional to the imaginary part of
the forward amplitude shown in fig.1.


The amplitude can be written as follows
\begin{equation}
  A = \frac{\kappa^2}{4\pi}
       \int~d^{D}Q~
       \frac{\Pi_{\mu\nu\rho\sigma}}{Q^{2}}
       \bra{n,\alpha;\tilde{n},\beta}
        V_{\mu\nu}(-Q)\Delta V_{\rho\sigma}(Q)
        \ket{n,\alpha;\tilde{n},\beta} \,,
\label{7*}
\end{equation}
where $\Pi_{\mu\nu\rho\sigma}(Q)=\eta_{\mu\nu} \eta_{\rho\sigma} + \dots$
is the polarization factor in the graviton propagator,
\[
   V_{\mu\nu}(Q) =
       \partial_{+}X^{\mu}\partial_{-}X^{\nu}
        \e^{-iQX}
\]
is the vertex operator for graviton, and $\Delta$ is the
string propagator. Hereafter the standard $i\epsilon$ prescription
is assumed in denominators. The evaluation of the matrix element
in eq.(\ref{7*}) is tedious but straightforward. One finds
\begin{eqnarray}
  A &=& \frac{\kappa^2}{4\pi}
      \xi^{\alpha}_{i}
      \xi^{\alpha}_{i'}
      \xi^{\beta}_{j}
      \xi^{\beta}_{j'}
       \int~d^{D}Q~
       \frac{\Pi_{\mu\nu\rho\sigma}}{Q^{2}}
       \int~dzd\bar{z}~
       B^{ii'}_{L,\mu\rho}(z)
       B^{jj'}_{R,\nu\sigma}(\bar{z}) \nonumber \\
&& \nonumber \\
&& \times~ z^{-\frac{1}{4}(PQ)-\frac{1}{2}(QL)+\frac{1}{8}Q^{2}-1}
       \left( 1-z \right)^{-\frac{1}{4}Q^{2}} \nonumber \\
&& \nonumber \\
&& \times~ (\bar{z})^{-\frac{1}{4}(PQ)+\frac{1}{2}(QL)
+\frac{1}{8}Q^{2}-1} \left( 1-\bar{z} \right)^{-\frac{1}{4}Q^{2}} \,,
\label{8+}
\end{eqnarray}
where
\begin{eqnarray} B^{ii'}_{L,\mu\rho}(z) &=& \frac{1}{n}
                      \int~\frac{du}{2\pi} \frac{du'}{2\pi}
                      \frac{1}{u^{n+1}} \frac{1}{(u')^{n+1}} \nonumber \\
&& \nonumber \\
&&\times~  \left( 1-u \right)^{-a} \left( 1-\frac{u}{z} \right)^{a}
           \left( 1-u' \right)^{a} \left( 1-zu' \right)^{-a}
           C^{ii'}_{\mu\rho}(z,u,u') \,. \label{8*}
\end{eqnarray} Here
\[
a= \frac{n+\tilde{n}}{2(e\cal{P})}(eQ) \,,
\]
and integration contours in complex $u$- and $u'$-planes are
small circles around the origin. The expression for $
C^{ii'}_{\mu\rho}$ is not particularly illuminating;
it is given in Appendix.

Expression similar to eq.(\ref{8*}) may be written also for
$ B^{jj'}_{R,\nu\sigma}(\bar{z})$.

The imaginary part of the amplitude emerges in the following way.
As is clear from eqs.(\ref{8*}) and (\ref{A5}), $B_{L}(z)$ entering
eq.(\ref{8+}) can be represented as finite sum (we omit indices
$i,i',\mu,\rho$ in what follows)
\[
    B_{L}(z) = \sum^{n}_{m=-n} B_{L,m} z^{-m}
\]
with real coefficients $B_{L,m}$. Similarly,
\[
    B_{R}(\bar{z}) = \sum^{\tilde{n}}_{\tilde{m}=-\tilde{n}}
                     B_{R,\tilde{m}} (\bar{z})^{-\tilde{m}} \,.
\]
At given $m$ and $\tilde{m}$, the integral over $z,\bar{z}$ is
straightforward to evaluate (one possibility is to make use of the
relation between closed and open string amplitudes \cite{GSW}).
One finds
\begin{eqnarray}
  A &=& \frac{\kappa^2}{4\pi}
      \xi\cdot
      \xi\cdot
      \xi\cdot
      \xi~
       \sum_{m,\tilde{m}}
       \int~d^{D}Q~
       \frac{\Pi}{Q^{2}}
       B_{L,m} B_{R,\tilde{m}}
       \sin \left( \frac{\pi Q^{2}}{4}\right) \nonumber \\
&& \nonumber \\
&& \times~  \frac{\Gamma\left(-\frac{1}{4}(PQ) - \frac{1}{2}(QL) - m
+ \frac{1}{8}Q^{2} \right)
         \Gamma\left( 1-\frac{1}{4}Q^{2} \right)}
       {\Gamma\left(-\frac{1}{4}(PQ) - \frac{1}{2}(QL) - m
         - \frac{1}{8}Q^{2} + 1 \right)} \nonumber \\
&& \nonumber \\
&& \times~  \frac{\Gamma\left(\frac{1}{4}(PQ) - \frac{1}{2}(QL) +\tilde{m}
        + \frac{1}{8}Q^{2} \right)
         \Gamma\left( 1-\frac{1}{4}Q^{2} \right)}
       {\Gamma\left(\frac{1}{4}(PQ) - \frac{1}{2}(QL) + \tilde{m}
         - \frac{1}{8}Q^{2} + 1 \right)} \,.
\label{10+}
\end{eqnarray}
The imaginary part of this integral is obviously due to the
region $Q^{2}\sim 0$ (graviton mass shell). Since the explicit factor
$Q^{2}$ from the graviton propagator is cancelled by
$ \sin \left( \frac{\pi Q^{2}}{4}\right)$, the imaginary part appears when
the product of gamma functions has double pole (Cutkosky rule).
This occurs when
\[
    \frac{1}{4}(PQ) + \frac{1}{2}(QL) + m = 0 \,,
\]
\begin{equation}
      \frac{1}{4}(PQ) - \frac{1}{2}(QL) +\tilde{m} = 0 \,.
\label{11*}
\end{equation}

Two remarks are in order. First, the compact component of the graviton
momentum is quantized in units $1/L$ \cite{GSW}, i.e.,
\[
     (QL) = r = integer.
\]
Equation (\ref{11*}) then implies
\begin{equation}
  \tilde{m} = m + r \,,
\label{11a*}
\end{equation}
i.e., $\tilde{m}$ is fixed for given $m$ and $Q$. Second, eq.(\ref{11*})
corresponds to the mass shell condition for the intermediate
macroscopic string state, as it should be.
Indeed, eqs.(\ref{11*}) and (\ref{5+}) imply
that the momentum of the intermediate state, $P'=P - Q$, obeys
the following relations,
\[
  - \frac{1}{4} (P')^{2} =
           (n -m) +(\tilde{n} - \tilde{m}) + \frac{1}{4}M_{0}^{2} \,,
\]
\begin{equation}
   (P'L) = (\tilde{n} - \tilde{m}) - (n -m) \,,
\label {11a+}
\end{equation}
which are precisely the mass shell conditions \cite{GSW} for physical
state of a string winding around the compact dimension, with mode
numbers $(n-m)$ and $(\tilde{n} - \tilde{m})$.

In the region (\ref{11*}), the integrand in eq.(\ref{10+}) becomes
\[
   -\frac{\pi}{4} B_{L,m} B_{R,\tilde{m}} ~~
 \frac{1}{\frac{1}{4}(PQ) + \frac{1}{2}(QL) + m - \frac{1}{8}Q^{2}}~~
 \frac{1}{\frac{1}{4}(PQ) - \frac{1}{2}(QL) + \tilde{m} +
                \frac{1}{8}Q^{2}} \,.
\]
To obtain the imaginary part, we recall the Cutkosky rule and obtain
\begin{equation}
    \mbox{Im} A = 4\pi^{2} \kappa^{2}
            \sum_{m} \int~d^{D}Q~ \Pi (Q)  B_{L,m} B_{R,\tilde{m}}
            \delta\left((PQ) + 2(QL) + 4m\right)
            \delta\left(Q^{2}\right) \,,
\label{11+}
\end{equation}
where $\tilde{m}$ is given by eq.(\ref{11a*}).

Let us now evaluate the dependence of $\mbox{Im} A$ on the length $L$ in
the regime (\ref{6*}). We begin with $B_{L,m}$. It has the following
representation (according to eq.(\ref{11a+}), positive $m$ are of
interest),
\[
    B_{L,m} = \frac{1}{2\pi i} \int~d\zeta~
         \frac{1}{\zeta^{m+1}} B_{L}\left(z=\frac{1}{\zeta}\right) \,,
   ~~~~~~~~~~~
\]
where the integration contour runs around the origin in complex
$\zeta$ plane. From eq.(\ref{8*}) we see that
\begin{eqnarray}
    B_{L,m} &=& \frac{1}{i}~\frac{1}{n} \int~\frac{d\zeta}{2\pi}
             ~\frac{du}{2\pi} ~\frac{du'}{2\pi}~
 \frac{1}{\zeta^{m+1}}~\frac{1}{u^{n+1}}~\frac{1}{(u')^{n+1}} \nonumber \\
&& \nonumber \\
&& \times~  \left( 1-u \right)^{-a}
            \left( 1-u\zeta \right)^{a}
            \left( 1-u' \right)^{a}
            \left( 1-\frac{u'}{\zeta} \right)^{-a}
             C\left(z=\frac{1}{\zeta},u,u'\right) \,.
\label{12*}
\end{eqnarray}
Now, recall that
\[
    n \propto L
\]
and also
\[
    m \propto L
\]
(the latter relation, valid for $Q$ independent of $L$,
 follows from eq.(\ref{11*}) and $P\propto L$). Therefore, we may use the
following asymptotic estimate
\newpage
\[
  \int~dz_{1}\dots dz_{k} z_{1}^{-\lambda a_{1}}\dots
              z_{k}^{-\lambda a_{k}}
               \Pi (1-z_{p})^{\alpha_{p}}
               \Pi (1-z_{p}z_{q})^{\beta_{pq}}
               \Pi (1-z_{p}/z_{q})^{\gamma_{pq}}~
\]
\begin{equation}
               \propto~
               \left(\frac{1}{\lambda}\right)^{\sum \alpha_{p} +
                 \sum \beta_{pq} + \sum \gamma_{pq} + k} \,,
\label{13*}
\end{equation}
which is valid as $\lambda \to  \infty$ with $a_{p},\alpha_{p},
\beta_{pq}, \gamma_{pq}$ fixed. This estimate is similar
to  ones used for evaluation of multi-Regge behavior of dual amplitudes,
and can be obtained in a way parallel to that
of ref. \cite{Dualamplitudes}. Applying this estimate to the integral
in eq.(\ref{12*}) we see that for obtaining the $L$-dependence of
$B_{L,m}$ we have only to count factors like
\begin{equation}
  \frac{1}{1-u},~~~\frac{1}{1-u'},~~~\frac{1}{1-uu'},~~~
  \frac{1}{1-u/z},~~~\frac{1}{1-zu'} \,,
\label{13a*}
\end{equation}
as well as explicit $L$-dependent factors, in $C(z,u,u')$: each factor
of the form (\ref{13a*}) and each factor $P$ produces a factor $L$,
while factors like $Q^{\mu},z,u,u'$ do not matter for the asymptotic
behavior of $B_{L,m}$. Making use of eq.(\ref{A5}) we obtain
\[
    B_{L,m} \propto \frac{1}{n}\left(\frac{1}{L}\right)^{3} L^{4} \,,
\]
where the first factor is explicit in eq.(\ref{12*}), the second factor
is due to three integrations ($k=3$ in the notation of eq.(\ref{13*})),
and the last factor is obtained by counting factors (\ref{13a*}) and
explicit
$L$-dependent factors in $C(z,u,u')$. So, we find that $B_{L,m}$ is
independent of $L$ in the limit of large $L$.

The remaining factors of $L$ in $\mbox{Im} A$, eq.(\ref{11+}),
come from two
sources. First, we have a factor $L^{-1}$ from
\[
   \delta\left((PQ) + 2(QL) + 4m\right)\propto
       \frac{1}{L} \delta\left(Q^0 - Q^1 - \frac{2m}{L}\right)
\]
(because $P=(2L,0,\dots,0) + O(1)$). Second, summation over $m$
gives a factor $L$,
\[
   \sum_{m} = L\int~d\left(\frac{m}{L}\right)
\]
(in other words, density of states of the final macroscopic
string is of order $L$). Combining these two factors, we see that
\[
   \mbox{Im} A =~~ independent~~ of~~ L \,.
\]

Finally, the graph of fig.1 represents the correction to $(mass)^2$
of the macroscopic string, and the decay rate $\Gamma$ is related
to $\mbox{Im} A$ as follows,
\[
   \Gamma = \mbox{Im}~ (mass) = \frac{1}{2M} \mbox{Im}~ (mass)^{2}
               \propto \frac {1}{L} \mbox{Im} A \,.
\]
So, we find{\footnote {Our estimate of  $B_{L,m}$ and $B_{R,\tilde{m}}$,
on which eq.(\ref{14*}) is based, would not be valid if there
were cancellations between contributions of various terms in $C(z,u,u')$.
In  that case the rate would be of order $L^{-2}$ or smaller.
However, the results of I imply that the partial decay rate into {\em some}
final states of macroscopic string is of order $L^{-1}(\ln L)^{-1}$,
so the total decay rate is at least of this order. This excludes the
dangerous cancellations and ensures that eq.(\ref{14*}) is indeed
correct.}}
\begin{equation}
    \Gamma \propto \frac{1}{L} \,.
 \label{14*}
\end{equation}

This behavior of the decay rate is precisely what one expects from the
point   of view of $(1+1)$ dimensions. Indeed, the state (\ref{7*}) is
normalized to contain two ``particles'' in the $(1+1)d$ universe of size
$L$. The interaction rate of these particles is of order $1/L$, in
agreement with eq.(\ref{14*}). We conclude that the emission of
a baby universe induced by interactions of ``particles'' has no
intrinsic suppression by the size of the parent universe.

\vskip3mm

{\bf 4.} String model of $(1+1)$-dimensional large and microscopic
universes provides an example of a theory where branching off of baby
universes induces the loss of quantum coherence in the parent universe.
The results of this paper show that the rate of this loss of coherence
is unsuppressed by the size of parent universes.
It remains to be understood whether this property is a peculiarity of
the string model, or it is generic to all theories allowing for
wormholes/baby universes.

\vskip7mm

{\bf Acknowledgements.}

\vskip1mm
The authors are indebted to M.Iofa, A.Isaev, G.Pivovarov and
P.Tinyakov for helpful discussions.
This work is supported in part by International Science Foundation
grant $\#$MKT300 and INTAS grant $\#$94-2352.

\vskip7mm

{\bf \large Appendix. }

\vskip3mm
For completeness, we present here the amplitude shown in fig.2.


 The initial and final states of the macroscopic string are
 DDF states with two ``particles'' in each of them;
 the momenta of the initial and final smooth strings,
 ${\cal P}$ and ${\cal P}'$ and the sets of vectors
$(e_{\mu}, {\bm \xi}_{\alpha})$ and
$(e'_{\mu}, {\bm \xi}'_{\alpha})$, used for the construction
of the DDF operators, need not be the same for initial
and final states.
The amplitude is written as follows,
\begin{eqnarray}
 A &=&\frac{\kappa^{2}}{4\pi}
      \zeta_{\mu\nu}(Q)\zeta_{\rho\sigma}(K) \nonumber \\
&& \nonumber \\
 && \times~  \biggl(
    \bra{n',\alpha';\tilde{n'},\beta'}
        V_{\mu\nu}(-Q)\Delta V_{\rho\sigma}(K)
        \ket{n,\alpha;\tilde{n},\beta}   \nonumber \\
&& \nonumber \\
&&+\, \bra{n',\alpha';\tilde{n'},\beta'}
        V_{\rho\sigma}(K)\Delta V_{\mu\nu}(-Q)
        \ket{n,\alpha;\tilde{n},\beta}
       \biggr) \,,
\label{A1}
\end{eqnarray}
where $\zeta (K)$ and $\zeta (Q)$ are polarizations of the
incoming and outgoing gravitons, respectively. The amplitude shown in
fig.1 is obtained from eq.(\ref{A1}) by setting
\begin{equation}
  K=Q,~~~ n=n',  ~~~ \tilde{n} =\tilde{n'},~~~ {\cal P} ={\cal  P}',~~~
   e=e',~~~ {\bm  \xi}={\bm \xi}',
\label{A2}
\end{equation}
substituting $\zeta\cdot\zeta$ by the graviton propagator
and integrating
over $Q$.

The evaluation of the amplitude (\ref{A1}) is straightforward. One finds
\begin{eqnarray}
 A &=& \frac{\kappa^{2}}{4\pi}
      \zeta_{\mu\nu}(Q)\zeta_{\rho\sigma}(K)
      \xi^{\alpha}_{i}
      \xi'^{\alpha'}_{i'}
      \xi^{\beta}_{j}
      \xi'^{\beta'}_{j'}
          \int~dzd\bar{z}~
       B^{ii'}_{L,\mu\rho}(z)
       B^{jj'}_{R,\nu\sigma}(\bar{z}) \nonumber \\
&& \nonumber \\
&& \times~ z^{-\frac{1}{4}(PK)-\frac{1}{2}(KL)+\frac{1}{8}K^{2}-1}
       \left( 1-z \right)^{-\frac{1}{4}(KQ)} \nonumber \\
&& \nonumber \\
&& \times~ (\bar{z})^{-\frac{1}{4}(PK)+\frac{1}{2}(KL)+\frac{1}{8}K^{2}-1}
\left( 1-\bar{z} \right)^{-\frac{1}{4}(KQ)} \,,
\label{A3}
\end{eqnarray}
 where
\begin{eqnarray}
B^{ii'}_{L,\mu\rho}(z) &=&
                    \frac{1}{\sqrt{nn'}} \int~\frac{du}{2\pi}
                    \frac{du'}{2\pi} \frac{1}{u^{n+1}}
\frac{1}{(u')^{n'+1}} \nonumber \\
&& \nonumber \\
&&  \times~ \left( 1-u \right)^{-a}
                     \left( 1-\frac{u}{z} \right)^{b} \left( 1-u'
                     \right)^{a'} \left( 1-zu' \right)^{-b'} \nonumber \\
&& \nonumber \\
&& \times~  \left(1-uu'\right)^{-c} C^{ii'}_{\mu\rho}(z,u,u')
\label{A4}
\end{eqnarray}
with
\[
a =
\frac{n+\tilde{n}}{2(e\cal{P})}(eQ) \,,~~~~~~
b =
    \frac{n+\tilde{n}}{2(e\cal{P})}(eK) \,,
\]
\[
a'=
\frac{n'+\tilde{n}'}{2(e'\cal{P}')}(e'Q) \,,~~~~~~~
b'=
    \frac{n'+\tilde{n}'}{2(e'\cal{P}')}(e'K) \,,
\]
\[
c =
    \frac{n+\tilde{n}}{(e\cal{P})}~~
      \frac{n'+\tilde{n}'}{(e'\cal{P}')}(ee') \,.
\]
The factor $ C^{ii'}_{\mu\rho}$ has the following form
\begin{eqnarray}
C^{ii'}_{\mu\rho}(z,u,u') &=&
   D_{1}^{i}D_{2}^{i'}R_{1}^{\mu}R_{2}^{\rho} +
   D_{1}^{i}R_{2}^{\rho}A_{1}^{i'\mu} +
   D_{1}^{i}R_{1}^{\mu}A_{2}^{i'\rho} +
   R_{1}^{\mu}R_{2}^{\rho}F^{ii'} \nonumber \\
&& \nonumber \\
&& +\, D_{2}^{i'}R_{1}^{\mu}B_{1}^{i\rho} +
   D_{2}^{i'}R_{2}^{\rho}B_{2}^{i\mu} +
   D_{1}^{i}D_{2}^{i'}G^{\mu\rho} \nonumber \\
&& \nonumber \\
&& +\, A_{1}^{i'\mu}B_{1}^{i\rho} +
   A_{2}^{i'\rho}B_{2}^{i\mu} +
   F^{ii'}G^{\mu\rho} \,,
\label{A5}
\end{eqnarray}
where
\[
  D_{1}^{i} =\frac{1}{2} {\cal P}_{L}^{i} -
             \frac{1}{2} Q^{i}~ \frac{u}{1-u}  +
             \frac{1}{2} K^{i}~ \frac{u/z}{1-u/z} -
             \frac{2n'e'^{i}}{(e'{\cal P}'_{L})}~\frac{uu'}{1-uu'} \,,
\]
\[
  D_{2}^{i'} =\frac{1}{2} {{\cal P}'}_{L}^{i'} +
             \frac{1}{2} Q^{i'}~ \frac{u'}{1-u'}  -
             \frac{1}{2} K^{i'}~ \frac{u'z}{1-u'z} -
             \frac{2ne^{i'}}{(e{\cal P}_{L})}~\frac{uu'}{1-uu'} \,,
\]
\[
  R_{1}^{\mu} = \frac{1}{4}\left(  P_{L}^{\mu} + {P'}_{L}^{\mu}\right)
               - \frac{n + \tilde{n}}{2(e{\cal P})}e^{\mu} \frac{1+u}{1-u}
               - \frac{n' + \tilde{n}'}{2(e'{\cal P}')}e'^{\mu}
               \frac{1+u'}{1-u'} - \frac{1}{2} K^{\mu} \frac{z}{1-z} \,,
\]
\[
  R_{2}^{\rho} = \frac{1}{4}\left( P_{L}^{\rho} + {P'}_{L}^{\rho}\right)
                - \frac{n + \tilde{n}}{2(e{\cal P})}e^{\rho}
                                    \frac{1+u/z}{1-u/z}
                - \frac{n' + \tilde{n}'}{2(e'{\cal P}')}e'^{\rho}
                                    \frac{1+u'z}{1-u'z}
                - \frac{1}{2} Q^{\rho} \frac{z}{1-z} \,,
\]
\[
  A_{1}^{i\mu} = \eta^{i\mu}\frac{u'}{\left(1-u'\right)^{2}}~,~~~~
  A_{2}^{i\mu} = \eta^{i\mu}\frac{u'z}{\left(1-u'z\right)^{2}} \,,
\]
\[
  B_{1}^{i\mu} = \eta^{i\mu}\frac{u/z}{\left(1-u/z \right)^{2}}~,~~~~
  B_{2}^{i\mu} = \eta^{i\mu}\frac{u}{\left(1-u \right)^{2}} \,,
\]
\[
  F^{ii'} = \delta^{ii'} \frac{uu'}{\left(1-uu' \right)^{2}}~,~~~~~
  G^{\mu\rho}= \eta^{\mu\rho} \frac{z}{\left(1-z \right)^{2}}~.
\]
The expression for the right factor $B_{R}(\bar{z})$ is completely
analogous to eq.(\ref{A4}).

The expression for the amplitude of fig.1, eq.(\ref{8+}) follows
from eq.(\ref{A3}). The factor $ C^{ii'}_{\mu\rho}$ entering
eq.(\ref{8*}) is obtained from eq.(\ref{A5}) by using the relations
(\ref{A2}).

\newpage
\unitlength=0.8mm
\special{em:linewidth 0.8pt}
\linethickness{0.8pt}
\begin{picture}(124.00,126.00)
\bezier{156}(104.00,97.00)(116.00,81.00)(104.00,66.00)
\bezier{16}(101.00,97.00)(103.00,98.00)(104.00,97.00)
\bezier{16}(101.00,66.00)(103.00,65.00)(104.00,66.00)
\bezier{156}(39.00,97.00)(51.00,81.00)(39.00,66.00)
\bezier{152}(36.00,97.00)(25.00,82.00)(36.00,66.00)
\bezier{16}(36.00,97.00)(38.00,98.00)(39.00,97.00)
\bezier{16}(36.00,66.00)(38.00,65.00)(39.00,66.00)
\put(57.00,91.00){\circle*{1.50}}
\put(92.00,91.00){\circle*{1.50}}
\bezier{24}(57.00,91.00)(55.00,94.00)(57.00,95.00)
\bezier{24}(57.00,95.00)(60.00,95.00)(59.00,98.00)
\bezier{20}(59.00,98.00)(59.00,101.00)(61.00,101.00)
\bezier{20}(61.00,101.00)(63.00,101.00)(63.00,104.00)
\bezier{24}(63.00,103.00)(63.00,107.00)(65.00,107.00)
\bezier{24}(65.00,107.00)(68.00,107.00)(68.00,110.00)
\bezier{24}(68.00,109.00)(68.00,112.00)(71.00,111.00)
\bezier{16}(71.00,111.00)(73.00,110.00)(75.00,111.00)
\bezier{20}(75.00,111.00)(78.00,112.00)(79.00,110.00)
\put(73.00,115.00){\line(0,-1){10.00}}
\put(73.00,101.00){\line(0,-1){10.00}}
\put(73.00,86.00){\line(0,-1){10.00}}
\put(73.00,71.00){\line(0,-1){10.00}}
\put(73.00,56.00){\line(0,-1){7.00}}
\put(73.00,120.00){\line(0,1){6.00}}
\put(84.00,115.00){\vector(2,-3){5.50}}
\put(58.00,106.00){\vector(2,3){5.50}}
\put(57.00,113.00){\makebox(0,0)[cc]{Q}}
\put(91.00,113.00){\makebox(0,0)[cc]{Q}}
\put(113.00,70.00){\vector(1,0){11.00}}
\put(11.00,70.00){\vector(1,0){11.00}}
\put(19.00,78.00){\makebox(0,0)[cb]{P}}
\put(117.00,77.00){\makebox(0,0)[cb]{P}}
\put(73.00,34.00){\makebox(0,0)[cc]{Fig. 1}}
\put(103.00,65.50){\line(-1,0){65.00}}
\put(38.00,97.50){\line(1,0){65.00}}
\bezier{20}(99.00,93.00)(98.00,91.00)(97.00,88.00)
\bezier{28}(100.00,69.00)(98.00,72.00)(97.00,75.00)
\bezier{24}(96.00,84.00)(95.00,81.00)(96.00,78.00)
\bezier{20}(81.00,107.00)(79.00,107.00)(79.00,110.00)
\bezier{28}(82.00,107.00)(85.00,107.00)(85.00,103.00)
\bezier{24}(85.00,103.00)(85.00,100.00)(88.00,100.00)
\bezier{16}(91.00,97.00)(90.00,95.00)(91.00,94.00)
\bezier{16}(91.00,94.00)(93.00,93.00)(92.00,91.00)
\bezier{20}(89.00,100.00)(91.00,100.00)(91.00,97.00)
\put(81.00,107.00){\line(1,0){1.00}}
\put(87.50,100.00){\line(1,0){1.50}}
\end{picture}

\unitlength=0.8mm
\special{em:linewidth 0.8pt}
\linethickness{0.8pt}
\begin{picture}(124.00,121.00)(0,20)
\bezier{156}(104.00,97.00)(116.00,81.00)(104.00,66.00)
\bezier{16}(101.00,97.00)(103.00,98.00)(104.00,97.00)
\bezier{16}(101.00,66.00)(103.00,65.00)(104.00,66.00)
\bezier{156}(39.00,97.00)(51.00,81.00)(39.00,66.00)
\bezier{152}(36.00,97.00)(25.00,82.00)(36.00,66.00)
\bezier{16}(36.00,97.00)(38.00,98.00)(39.00,97.00)
\bezier{16}(36.00,66.00)(38.00,65.00)(39.00,66.00)
\put(57.00,91.00){\circle*{1.50}}
\put(92.00,91.00){\circle*{1.50}}
\put(113.00,70.00){\vector(1,0){11.00}}
\put(11.00,70.00){\vector(1,0){11.00}}
\put(19.00,78.00){\makebox(0,0)[cb]{P}}
\put(117.00,77.00){\makebox(0,0)[cb]{P'}}
\put(71.00,46.00){\makebox(0,0)[cc]{Fig. 2}}
\put(103.00,65.50){\line(-1,0){65.00}}
\put(38.00,97.50){\line(1,0){65.00}}
\bezier{20}(99.00,93.00)(98.00,91.00)(97.00,88.00)
\bezier{28}(100.00,69.00)(98.00,72.00)(97.00,75.00)
\bezier{24}(96.00,84.00)(95.00,81.00)(96.00,78.00)
\bezier{28}(92.00,91.00)(95.00,93.00)(92.00,95.00)
\bezier{28}(92.00,95.00)(89.00,97.00)(92.00,99.00)
\bezier{28}(92.00,99.00)(95.00,101.00)(92.00,103.00)
\bezier{24}(92.00,103.00)(90.00,105.00)(92.00,107.00)
\bezier{24}(92.00,113.00)(90.00,115.00)(92.00,117.00)
\bezier{28}(92.00,117.00)(95.00,119.00)(92.00,121.00)
\put(57.00,91.00){\circle*{1.50}}
\bezier{28}(57.00,91.00)(60.00,93.00)(57.00,95.00)
\bezier{28}(57.00,95.00)(54.00,97.00)(57.00,99.00)
\bezier{28}(57.00,99.00)(60.00,101.00)(57.00,103.00)
\bezier{24}(57.00,103.00)(55.00,105.00)(57.00,107.00)
\bezier{24}(57.00,113.00)(55.00,115.00)(57.00,117.00)
\bezier{28}(57.00,117.00)(60.00,119.00)(57.00,121.00)
\put(52.00,105.00){\vector(0,1){13.00}}
\put(98.00,118.00){\vector(0,-1){12.00}}
\put(45.00,112.00){\makebox(0,0)[cc]{Q}}
\put(106.00,112.00){\makebox(0,0)[cc]{K}}
\bezier{32}(92.00,107.00)(95.00,110.00)(92.00,113.00)
\bezier{32}(57.00,107.00)(60.00,110.00)(57.00,113.00)
\end{picture}

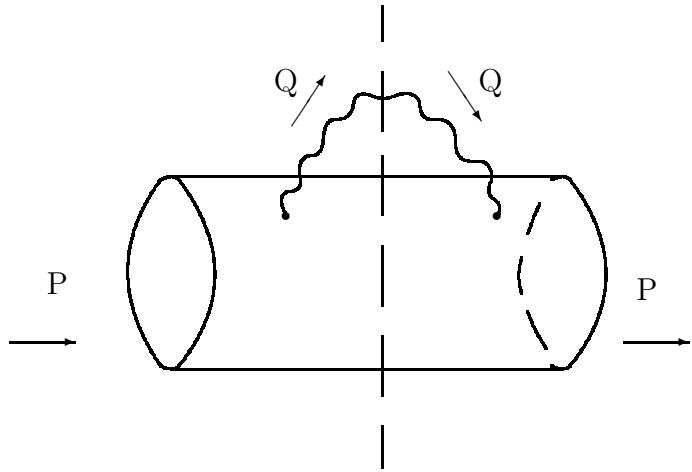
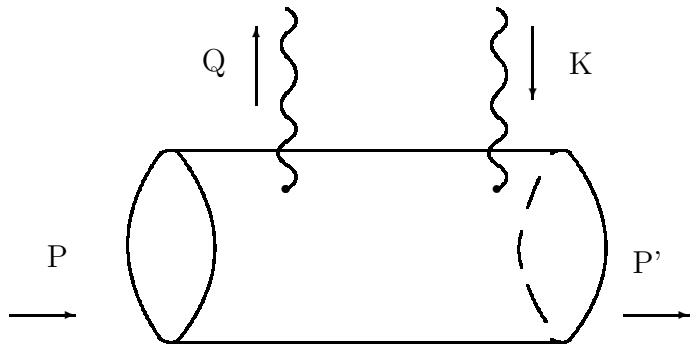
\end{document}